\documentclass[conference]{IEEEconf}
\IEEEoverridecommandlockouts
\usepackage{cite}
\usepackage{amsmath,amssymb,amsfonts}
\usepackage{gensymb}
\usepackage{algorithmic}
\usepackage{graphicx}
\usepackage{textcomp}
\usepackage{xcolor}
\usepackage{tabularx}
\usepackage[hidelinks]{hyperref}

\usepackage{booktabs}

\renewcommand{\baselinestretch}{0.92}

\def\BibTeX{{\rm B\kern-.05em{\sc i\kern-.025em b}\kern-.08em
    T\kern-.1667em\lower.7ex\hbox{E}\kern-.125emX}}
\usepackage{graphicx} %

\title{\LARGE
Sensor-Placement-Agnostic Sonomyography: Toward Continuous High-Dimensional Control by Users with Tetraplegia
}
\author{Gavin Sueltz, Vikram Athithan*, Emma Ferran*, Maria Herrera*, Carson J. Wynn*, and Laura A. Hallock
\thanks{*These authors contributed equally to this work.}
\thanks{This work was supported by the University of Utah Research Fund, as well as the Department of Mechanical Engineering and the Office of Undergraduate Research at the University of Utah.}
\thanks{The authors are with the Departments of Mechanical Engineering and Physical Medicine \& Rehabilitation, the Kahlert School of Computing, and the Robotics Center at the University of Utah, Salt Lake City, UT 84112, USA.
Correspondence should be directed to {\tt\footnotesize laura.hallock@utah.edu}.}
\thanks{The study protocol employed in this work was approved by the University of Utah Institutional Review Board under Protocol IRB\_00183389, and written informed consent was obtained from each study participant.}%
}

\begin{document}

\bstctlcite{IEEEexample:BSTcontrol}

\maketitle

\begin{abstract}
Sonomyography (SMG) enables continuous device control via ultrasound-measured muscle deformation signals, but existing SMG interfaces generally require substantial user- and sensor-location-specific training data and provide only one proportional signal or task-specific classification.
We present a real-time, sensor-placement-agnostic SMG control system based on 
sparse optical flow tracking that enables continuous 1-DOF
control after minimal calibration (3 pose definitions). We also present a preliminary expansion of this method that augments this algorithm with a short computer-aided calibration to enable 2-DOF control.

We evaluate both 1- and 2-DOF systems' performance for a preliminary cohort of 
3 cervical spinal cord injury survivors and 6 uninjured individuals across 6 sensor placements spanning the arm, neck, and upper torso.
As assessed by a cursor trajectory tracking task, all participants achieved continuous 1-DOF control at all tested sensor locations (even those that relied on passive tissue motions), with all participants achieving $<$5.5\% tracking error using at least one placement (and many $<$4\% across many). 
All participants were also able to modulate 2D cursor position via the 2-DOF system, with varying levels of control authority, 
and several were able to complete a 2D drawing task,
constituting the first (to our knowledge) demonstration of location-agnostic multi-DOF continuous SMG-based control.
These results highlight the promise of SMG to enable rapidly calibratable, high-dimensional, sensor-placement-agnostic device control by users with tetraplegia, and also illuminate key challenges in both signal processing and practical system deployment. To enable further development by scientific and user communities,
developed algorithms have been open-sourced
as part of the OpenMyoControl project on SimTK ({\tt \href{https://simtk.org/projects/openmyocontrol}{simtk.org/projects/openmyocontrol}}).
\end{abstract}

\section{Introduction}

    Spinal cord injury (SCI) survivors and other individuals with tetraplegia could benefit from continuous, task-agnostic control interfaces for robot manipulators and other high-degree-of-freedom (DOF) assistive devices. However, commercially available interfaces (joysticks~\cite{henderson2013review, rulik2022control, aspelund2020controlling}, sip-and-puff~\cite{mougharbel2013comparative}, etc.) generally provide only low-DOF signals that map poorly to such devices. Biosensing interfaces --- surface electromyography (sEMG), sonomyography (SMG)~\cite{engdahl_sonomyography_2024}, and others --- are a promising mechanism to provide 
    more intuitive and accessible control signals.
    Existing systems, however, require substantial user- and sensor-location-specific training data, and all but the most data-intensive systems are unable to provide more than a single continuous control signal or task-specific classification, 
    limiting utility for tetraplegic users with heterogeneous residual function and assistance requirements and low tolerance for long calibration times. 

    In this work, we address both the need for sensor-location-agnostic control systems that leverage SCI survivors' heterogeneous residual function and the need for systems supporting higher-dimensional control without the addition of (expensive) additional hardware or long calibration times. Specifically, building on our prior investigations of optical-flow-based SMG signals~\cite{hallock2020muscle, hallock_toward_2021}, 
    we present the following contributions:

\begin{figure}[!t]
    \centering
    \includegraphics[width=0.8\linewidth]{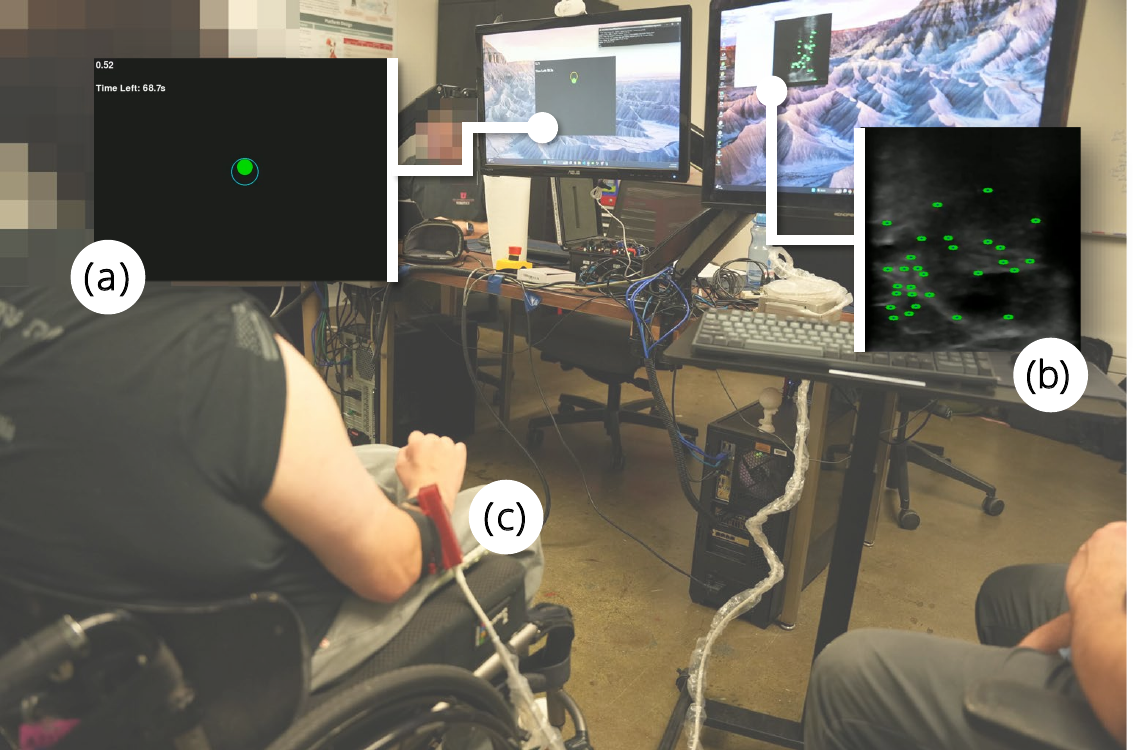}
    \caption{
    While biosensing interfaces are a promising tool with which to measure high-dimensional signals to allow users with tetraplegia to control assistive devices, existing systems largely require substantial user- and sensor-configuration-specific training data and calibration and provide only one continuous intent signal or task classification. In this work, we present and evaluate a novel optical-flow-based sonomyography (SMG) control method enabling users to control the vertical position of on-screen cursor (a) using SMG data (b) from probe (c) placed on arbitrary body locations after minimal calibration (3 pose definitions). In addition, we present (to our knowledge) the first demonstration of continuous location-agnostic 2D SMG-based control via a 2-DOF expansion to this 1-DOF algorithm.}
    \vspace{-1em}
    \label{fig:system}
\end{figure}

    \begin{itemize}
        \item development of a novel optical-flow-based SMG control method (the usage of which is depicted in Figure~\ref{fig:system}) that enables continuous 1-DOF signal extraction from B-mode ultrasound data collected from arbitrary sensor locations with rapid 3-point calibration;
        \item expansion of this algorithm to permit the extraction of two independent control signals from a single probe, enabling proof-of-concept 2-DOF control;
        \item evidence
        that the 1-DOF system is performant and well-received when used by a pilot cohort of cervical SCI survivors and uninjured users
        with a wide range of sensor placements, and that 2-DOF system usage is feasible in this population; and
        \item a preliminary characterization of 1- and 2-DOF system performance and user preferences, evaluated across injury levels, sensor placements, and selected motions,
        informing directions for future system extension and refinement. 
    \end{itemize}
To enable expansion on this work, and early access to such algorithms by assistive device users, 1- and 2-DOF algorithms 
have been made available as part of the OpenMyoControl project on SimTK ({\tt \href{https://simtk.org/projects/openmyocontrol}{simtk.org/projects/openmyocontrol}}), alongside our collected data.

\section{Background and Related Work}

Recovery of high-dimensional control capabilities like those of the healthy hand and arm is a primary priority~\cite{anderson2004targeting} for the hundreds of thousands of individuals with tetraplegia due to cervical SCI~\cite{center2024traumatic}
and other disorders in the United States alone. Even SCI survivors with substantial remaining volitional control of the arms generally have substantial impairment at the hands,
such that most tetraplegic users struggle to perform tasks like door opening, object pickup, and glasses adjustment that could be addressed by control of an assistive manipulator. Most available interfaces are poorly suited to control of devices more complex than wheelchairs (e.g., multi-step, move-and-click joystick interfaces~\cite{kinova2021jacoug}), and/or targeted at users with no remaining arm function (e.g., sip-and-puff~\cite{mougharbel2013comparative} and other mouth interfaces), failing to take advantage of many tetraplegic users' residual capabilities and resulting in uneven technology adoption.

Biosensing interfaces --- commonly, sEMG, but also emerging modalities like SMG --- are a promising source of high-dimensional signals to drive such systems and can be collected from different locations on the body at which individual users retain function.
At the same time, while such systems have demonstrated impressive performance in multi-DOF intent inference for highly specific goal motions and sensor locations (e.g., emg2pose~\cite{salter_emg2pose_2024} at the wrist), these solutions generally require extensive training data collection and calibration procedures, limiting generalizability to new user cohorts and motion tasks.
Additionally, for sEMG-based interfaces, the nature of the human neurological system --- consisting of probabilistic spiking of discrete neural signals --- makes it inherently challenging to convert these signals into continuous, task-suitable values readily manipulable by users.

Sonomyography (SMG) is an alternative (complementary) biosensing modality 
    that avoids this processing challenge by measuring smooth mechanical outputs of the chaotic neurological system (i.e., muscle and tissue deformations), also from anywhere on the body,  resulting in continuous signals~\cite{dhawan_proprioceptive_2019}.
    There are a number of methods to parameterize the SMG signal~\cite{song_review_2023}, and several groups have demonstrated both single-DOF proportional control~\cite{shenbagam_sonomyography-based_2024} and discrete classification-based control with a high number of categories~\cite{dhawan_proprioceptive_2019}, supporting SMG signals' high-dimensional information content.
    However, as with sEMG, these systems require collection of substantial user- and sensor-location-specific training data 
    to predict intent, similarly limiting applicability in new populations with heterogeneous neuromusculoskeletal function.
    It also remains unclear how these systems (which often rely on full-image correlations~\cite{dhawan_proprioceptive_2019})
    can be expanded to perform multi-DOF continuous control without prohibitively high-dimensional training data collection (or the use of many ultrasound probes, which remain bulky and expensive).
    
    In prior work, we piloted an alternative SMG formulation in which Lucas--Kanade sparse optical flow~\cite{Lucas1981} was used to measure specific muscle contraction motions 
    over time in order to both assess muscle force--deformation dynamics~\cite{hallock2020muscle} and show proof-of-concept optical-flow-based SMG control~\cite{hallock_toward_2021}.
    In this paper, noting that not only muscles, but all tissues deform during such motions (providing expanded features to track over time), we introduce and evaluate a novel optical-flow-based SMG control formulation that is agnostic to both sensor placement and the specific tissues tracked, as long as the user is able to generate sufficient deviations in feature locations via self-selected motions. This approach allows the generation of continuous signals from an arbitrary sensor location with a minimal 3-pose-definition calibration procedure. 
    We also introduce a naive extension of this formulation to 2-DOF control, enabling multi-dimensional control without precise sensor placement requirements or additional probes. We evaluate a pilot cohort of SCI survivor and uninjured users' performance and preferences when utilizing both systems at a variety of sensor locations.
    Below, we present the details of these algorithms' formulations and 
    hardware requirements (section~\ref{sec:sysdes}), the scope and methods of our preliminary evaluation study (section~\ref{sec:syseval}), results and insights on participants' performance and preferences (section~\ref{sec:sysevalres}), and identified limitations and directions for future investigation (section~\ref{sec:conc}).

\section{System Design}\label{sec:sysdes}

\begin{figure*}[!t]
        \centering
        \includegraphics[width=\textwidth]{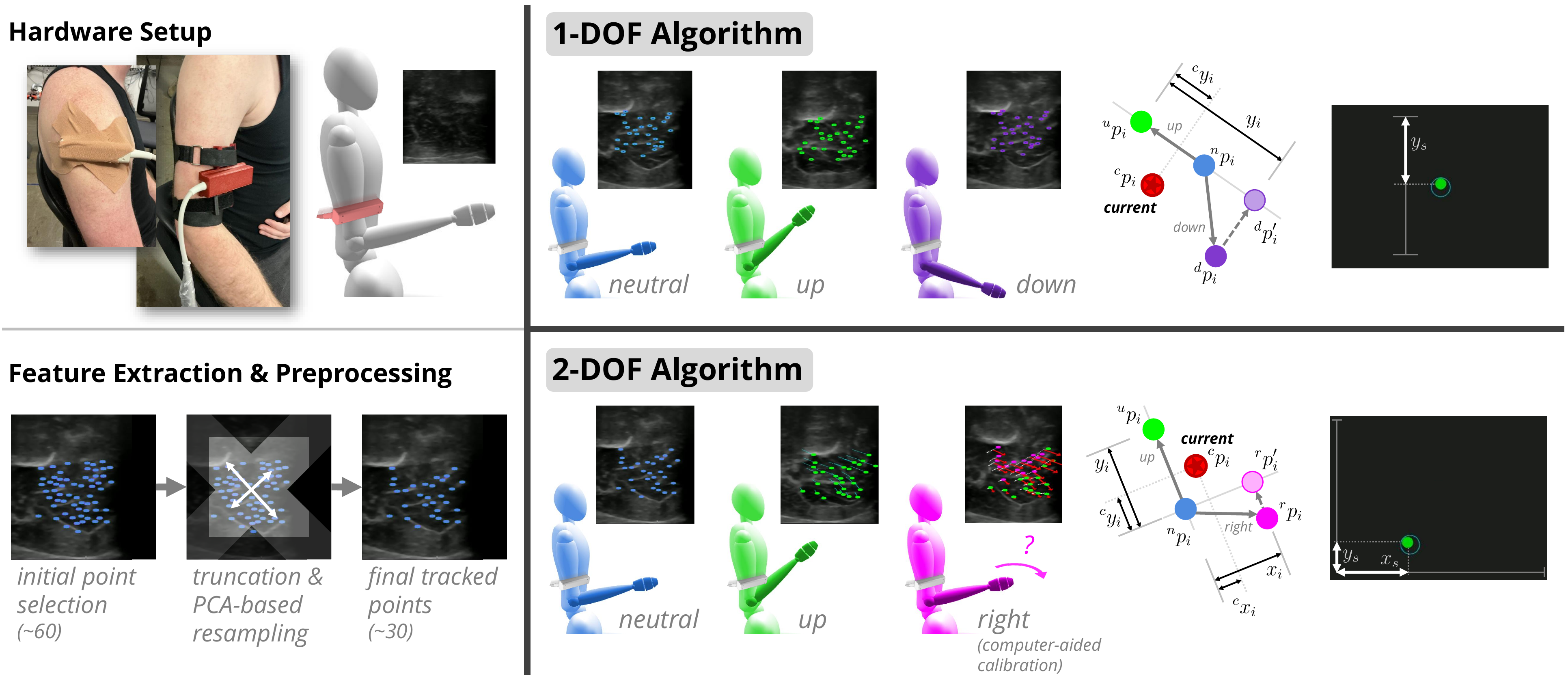}
        \caption{Hardware setup, signal preprocessing, and algorithmic details enabling 1- and 2-DOF cursor control from a single B-mode ultrasound time series collected from an arbitrary location on the body, depicted through biceps imaging modulated through arm motion. \textit{Left, Top}: Ultrasound probe is affixed to desired sensor location (e.g., deltoid, \textit{left}, or biceps, \textit{right}) using kinesio tape or a 3D printed mount and straps, as detailed in section~\ref{sec:hwset}. \textit{Left, Bottom}: Initial points to track are identified, then resampled along principal axes of variation (eliminating those close to image edges) to generate a refined set of points for tracking that are well-distributed throughout the image, as detailed in section~\ref{sec:pproc}. \textit{Right, Top}: 1-DOF control of the cursor (\textit{right}) is achieved by relating the aggregate positions of tracked points at user-selected ``neutral'' (\textit{blue}), ``up'' (\textit{green}), and ``down'' (\textit{purple}) poses, which are then linearly interpolated based on current point positions (\textit{red}) to generate a continuous signal, as detailed in section~\ref{sec:sys1dof}. \textit{Right, Bottom}: 2-DOF cursor control (\textit{right}) is achieved by first defining ``neutral'' and ``up'' poses as above, then defining a ``right'' pose at which a sufficient number of displayed points turn \textit{magenta}, indicating that associated right vector $^np_i$ to $^rp_i$ is sufficiently orthogonal to up vector $^np_i$ to $^up_i$. Horizontal and vertical cursor positions are then linearly interpolated in the same manner as the 1-DOF algorithm along each axis, as discussed in section~\ref{sec:sys2dof}. \textit{Note}: Some ultrasound images have been cropped to areas of interest and slightly altered 
        to aid clarity in our algorithm explanations. These changes are restricted to modifying point colors and adding additional tracked points to the first two processing pipeline steps to illustrate the process, as this step was not displayed graphically during system usage.
        }\vspace{-1em}
        \label{fig:meth}
    \end{figure*}

We present two novel algorithms for extracting continuous control signals from a single B-mode ultrasound time series, both illustrated in Figure~\ref{fig:meth}: a 1-DOF algorithm (the primary focus of our analysis) and a (preliminary) 2-DOF algorithm. Both algorithms use the same hardware setup and image preprocessing, as detailed below. Implementations of these algorithms are included with code release.

\subsection{Hardware Setup, Test Output System, \& Signal Requirements}\label{sec:hwset}

The algorithms presented were implemented in Python on an ArtUs EXT-2H Ultrasound system with linear probe LF9-5N60-A3 (Telemed Medical Systems, Milan, Italy) depicted in Figure~\ref{fig:meth} (left, top), but do not depend on any specific hardware except in details of network interfacing. Images were $12 \times 12$~cm in size (512 $\times$ 512~px) and collected at a rate of 30~Hz. As shown, probes were snugly attached to the body using either a 3D printed mount and straps or kinesiology tape depending on location and user preference.

The 1- and 2-DOF algorithms' outputs (1 and 2 continuous control signals, respectively) were mapped to movement of a cursor on a screen in 1 or 2 dimensions, respectively (Figure~\ref{fig:meth}, right). The presented algorithms are agnostic to sensor attachment location and selected motions defining cursor movement direction, with the following caveats. First, the image must contain a sufficient number of trackable features (see section~\ref{sec:pproc}). Second, the identified features must move a sufficient distance between selected motion extrema to enable defining a control workspace, and, in the 2-DOF case, the direction of feature motion between up/down and left/right motions must be sufficiently orthogonal, for a sufficient number of features, as defined more precisely below in section~\ref{sec:sys2dof}.\footnote{Empirically, in our pilot testing in section~\ref{sec:syseval}, functional attachment locations were largely easy to find across examined body areas except as otherwise noted and did not require precise placement. A more rigorous treatment of how to know whether a particular placement is suitable is a compelling direction for future research.}

\subsection{Feature Extraction \& Preprocessing}\label{sec:pproc}

Both 1- and 2-DOF algorithms utilize the standard iterative Lucas--Kanade method of sparse optical flow estimation~\cite{Lucas1981}, as implemented in the OpenCV Python library~\cite{opencv_library}, to track the motion of selected image points over time. Points used in tracking are selected over the multi-step, empirically-informed preprocessing pipeline depicted in Figure~\ref{fig:meth} (left, bottom) to ensure that chosen points are both distributed throughout the image (not overly clustered) and unlikely to leave the imaging window during tissue motion. Specifically, 60~points are initially selected on the basis of their Shi--Tomasi corner score~\cite{shi_good_1994}. Points within 20\% of the image border are then removed and principal component analysis performed on the remaining points to determine the primary direction along which the point cloud is distributed, as a proxy for directions along which promising tissue features can be located. A final set of up to 30~points is then sampled from bounded regions surrounding the principal component vectors. Further implementation details (including specific score thresholds used in point selection and sampling region definitions) can be found in the accompanying code release.

\subsection{1-DOF Algorithm}\label{sec:sys1dof}

Once the final set of points to track has been established through the procedures in section~\ref{sec:pproc}, users are asked to define a ``neutral'' pose, followed by poses corresponding to ``up'' and ``down'' extrema of the on-screen cursor. Figure~\ref{fig:meth} (right, top) depicts this process for a sensor placement on the biceps, with elbow flexion/extension as the controlling motion, but arbitrary placements and motions are possible (and many explored in section~\ref{sec:syseval}).

The control scheme for the cursor's up/down motion is then defined as follows, and as depicted in Figure~\ref{fig:meth}'s point diagram. For each of the $m$ tracked points $p_i$ (with corresponding neutral, up, and down locations $^np_i$, $^up_i$, and $^dp_i$), $^dp_i$ is first projected onto the line defined by $^np_i$ and $^up_i$ to generate collinear point $^dp_i '$. Defining $y_i$ as the total distance between $^up_i$ and $^dp_i '$, and $^cy_i$ as the distance along this line between $^up_i$ and current point $^cp_i$, vertical location $y_s$ of the on-screen cursor (measured from the top) is then defined as 
\begin{equation}\label{eq:1dof}
    y_s = \mathrm{clamp}\left( \frac{1}{m}\sum_{i=1}^m\frac{^cy_i}{y_i}\right)
\end{equation}
where
\begin{equation}
    \mathrm{clamp}(x) = \begin{cases} 0 & x < 0 \\ x & 0 \le x \le 1 \\ 1 & x > 1 \end{cases}
\end{equation}
i.e., the mean distance of all tracked points along their corresponding directions of variation, truncated at the defined up/down extrema.

\subsection{2-DOF Algorithm}\label{sec:sys2dof}
To enable cursor control in two dimensions, we follow the same steps as the 1-DOF algorithm through selection of ``neutral'' and ``up'' poses, then employ a computer-aided calibration method to select a ``right'' pose, as depicted in Figure~\ref{fig:meth} (right, bottom). Specifically, users are guided to select a ``right'' pose at which a sufficient number of points (at least 50\% of available points generally provides reasonable performance) deviate within 30$\degree$ of orthogonal to their ``up'' pose locations. This instruction is provided visually: users observe the ultrasound image and points satisfying this constraint turn magenta from their displayed green.

Up/down and left/right motions are then defined in an analogous method to the 1-DOF algorithm, this time with cursor locations $x_s$ (horizontal) and $y_s$ (vertical) measured from the bottom left corner. Distance $y_s$ is again calculated using Equation~\ref{eq:1dof}, with the only difference that $y_i$ is calculated as distance between $^up_i$ and $^np_i$, and $x_s$ in an identical manner for all orthogonal quantities.

\section{Preliminary System Evaluation: Methods}\label{sec:syseval}

We evaluated the section~\ref{sec:sysdes} systems in a preliminary cohort of SCI survivors and uninjured users to establish usability by individuals with tetraplegia and to gain initial insights into system performance across sensor locations and movement types, as well as user preferences. Study methodology is discussed below, with corresponding discussion of findings in section~\ref{sec:sysevalres}.

\subsection{Participant Demographics}

Data were collected from 9 participants: 3 cervical SCI survivors and 6 uninjured individuals. All participants were right-handed. 3 SCI participants (1, 2, and 9) were male and reported cervical SCI injury 1--7~y prior to the study. Participant 1, age 25, reported a complete SCI of the C5--7 vertebrae. He self-reported having more volitional control than most C5 complete SCI survivors, including enough arm control and sensation to use an unpowered wheelchair. Participant 2, age 59, reported a partial C5 SCI. He reported volitional control and sensation of the arms, but very little to none in the hands, and used a powered wheelchair. Participant 9, age 30, reported 2 SCIs: an old T4 complete, and a more recent C7 partial. He reported a lack of feeling in his shoulders, but had intact volitional control and sensation in his arms and hands and used an unpowered wheelchair.\footnote{We aim to recruit participants with more severe injuries (e.g., less volitional control below the shoulders) in future studies to provide more robust insights on the utility of this system across the spectrum of SCI injury.}

The uninjured cohort (participants 3--8) consisted of 3 male and 3 female participants, age $29.2 \pm 14.2$~y (range: 21--58~y). Midway through the study, participant 7 (female, age 58) disclosed adhesive capsulitis (``frozen shoulder'') but elected to continue, and her data is evaluated in this context.

\subsection{Consent \& Subjective Data Collection}

After providing written informed consent on arrival, each participant completed an intake survey to obtain their baseline affect using the Self-Assessment Manikin (SAM) \cite{bradleyMeasuringEmotionSelfAssessment1994}. The SAM was also completed after each assessed sensor placement alongside Likert-scale \cite{likert1932technique} and free-response assessments of comfort, control, fatigue, and preference across motion options. Notes on participant feedback when conversing during experiments were also recorded by investigators for inclusion in subjective analysis.

\subsection{1-DOF System Evaluation}

The 1-DOF algorithm detailed in section~\ref{sec:sys1dof} was evaluated by all participants across the 6 sensor placements outlined in Table~\ref{table:chosen}, in order, even those for which the participant reported no active volitional control (e.g., wrist flexors). For each placement, the ultrasound probe was attached to the user transverse to the indicated muscle(s) as documented in section~\ref{sec:hwset}. Placement was adjusted as needed by investigators based on real-time inspection of the ultrasound image to ensure sufficient adherence to section~\ref{sec:hwset}-defined signal qualities and minimal probe shifting, but were not rigorously located based on anatomical landmarks, as robustness to imprecise placement is a valuable aspect of the system.

For each placement, participants were instructed to calibrate the 1-DOF system's ``up'' and ``down'' poses in two ways, documented in Table~\ref{table:chosen}: first, using a motion prescribed by investigators, and second, using a self-selected motion. Motion magnitude was not prescribed, but users were guided toward smaller motions that utilized a comfortable range of motion that did not require large effort, approach joint limits, or result in tracked points moving out of the image.

Following each calibration, users were first allowed unbounded time to practice moving the cursor up and down the screen. If the user did not feel that control was adequate, the system was recalibrated a maximum of 3 times before characterizing that placement--movement combination as not viable for that user. When users required recalibration, it was generally due to improper sensor attachment on body locations with challenging geometry, most commonly the sternocleidomastoid (SCM) and the trapezius (TRA).
Once a participant elected to proceed, they were instructed to perform a 2~min trajectory tracking task (see Figure~\ref{fig:system}) of incrementally increasing difficulty by following a target circle as it traversed a stair-step move-and-hold, slow sine wave, fast sine wave, and arbitrary rapid transition pattern, as shown in Figure~\ref{fig:ErrVsTime}. This task (including both calibration and trajectory following) was performed twice for each sensor placement--movement combination while both target and actual cursor positions were recorded. Participants were instructed not to talk during SCM trials to avoid generating confounding neck tissue motions.

\begin{table}[t]
\centering
\small
\centering
\caption{Sensor Placements \& Motions (1-DOF Assessment)}
\label{table:chosen}
\begin{tabular}{
>{\raggedright\arraybackslash}p{0.03\textwidth}
>{\raggedright\arraybackslash}p{0.08\textwidth}
>{\raggedright\arraybackslash}p{0.08\textwidth}
>{\raggedright\arraybackslash}p{0.2\textwidth}
}
\toprule
\textbf{ID} & \textbf{Location} & \textbf{Prescribed Motion} & \textbf{Self-Selected Motion} \\
\midrule
BIC & biceps & elbow flexion / extension & forearm supination / pronation (1, 2, 4), shoulder internal / external rotation (3, 5--9) \\ 
\addlinespace
SCM & sternocleido-mastoid & right / left head tilt & head flexion / extension (1, 3--9), looking right / left (2) \\ 
\addlinespace
TRA & trapezius & shoulder elevation / depression & shoulder protraction / retraction (1, 2, 5, 7--9), shoulder abduction / adduction (3, 6), shoulder flexion / extension (4) \\ 
\addlinespace
DEL & deltoid & shoulder abduction / adduction & shoulder external / internal rotation (1, 4, 8), shoulder protraction / retraction (2), shoulder flexion / extension (3, 5--7, 9)\\ 
\addlinespace
EXT & wrist extensors & wrist flexion / extension & forearm supination / pronation (1--3, 7, 8), hand open / close (4--6, 9) \\  
\addlinespace
FLE & wrist flexors & wrist flexion / extension & forearm supination / pronation (1--3, 5--7, 9),  hand open / close (4, 8)\\  

\addlinespace
\bottomrule
\end{tabular}
\vspace{-1em}
\end{table}

\subsection{2-DOF System Evaluation}

To assess feasibility of the 2-DOF system while maintaining a tractable experiment duration (and in deference to the systems' preliminary nature), data were collected from a single self-selected sensor location from each participant. Participants were re-sensorized with the probe at their elected location and underwent the calibration procedure detailed in section~\ref{sec:sys2dof}. This procedure was challenging, and participants were allowed an unbounded number of recalibrations and sensor repositionings. As when evaluating the 1-DOF system, participants were allowed unbounded time to practice moving the cursor around the screen. Once they elected to proceed, they were instructed to perform two trajectory tracking tasks, again following a target circle, shown in Figure~\ref{fig:meth} (right): an out-and-back screen traversal task designed to assess participants' reachable workspace, and a drawing task of a stylized letter ``U'' (see Figure~\ref{fig:blockU}) to assess users' ability to follow a 2D trajectory by precisely modulating each signal independently. Participants were granted unlimited (in practice, several) attempts to complete both tasks.

\section{Preliminary System Evaluation: Results \& Discussion}\label{sec:sysevalres}

In this section, we report insights on 1- and 2-DOF system performance enabled by our section~\ref{sec:syseval} data collection, as quantified via users' trajectory tracking error and workspace traversal, in the context of reported user preferences, including implications for future system development and expansion.

\subsection{1-DOF System Performance}

Observations on the 1-DOF system's performance across participants, sensor locations, and task types are reported below.

\subsubsection{Error Metric: RMSE}\label{sec:metric}

To assess users' performance using a given 1-DOF control scheme, we use the aggregate root mean square error (RMSE), defined as the mean Euclidean distance (i.e., absolute value) between the desired (trajectory) position and the user-controlled cursor position over time. These RMSE values should be understood as the unitless fraction of the total traversable distance (i.e., the total height of the screen is 1).

\subsubsection{Performance Across Injury Levels \& Sensor Locations} %

\begin{figure}[!t]
        \centering
        \includegraphics[width=\linewidth]{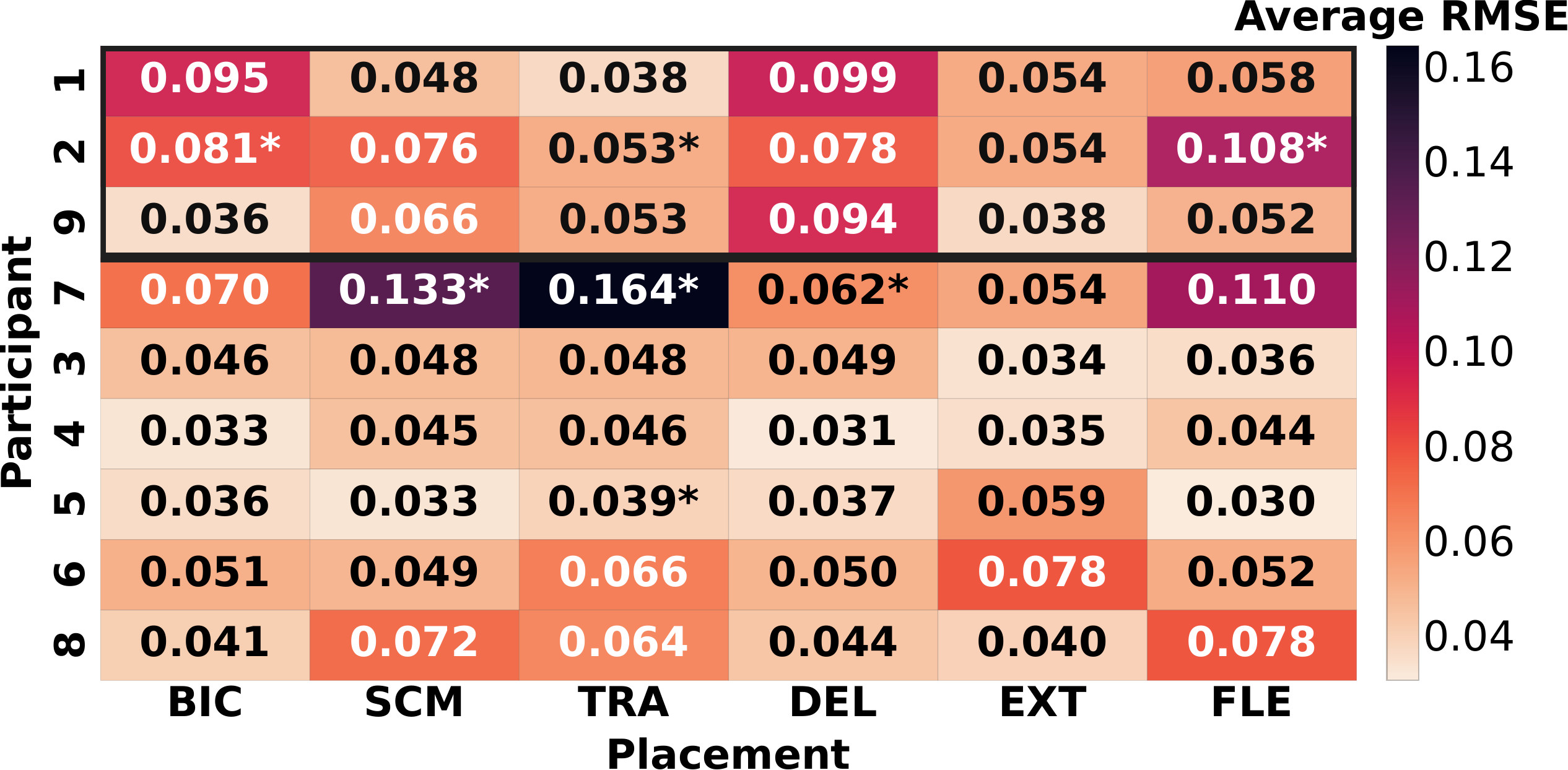}
        \caption{Average RMSE during the 1-DOF trajectory tracking task, calculated as noted in section~\ref{sec:metric}, for each participant and sensor placement, averaged across all successful trials for both prescribed and self-selected motions as listed in Table~\ref{table:chosen}. SCI survivor participants 1, 2, and 9 (\textit{top}, boxed), were able to complete most tasks with only slightly higher error than most uninjured participants, even when the probe was located along muscles at which the user had no volitional control (e.g., FLE for participant 2, whose wrist extensors generated sufficient flexor tissue motion). Participant 7 (who reported adhesive capsulitis) exhibits substantially higher error than others when using shoulder muscle control schemes, likely as a consequence of her condition. Asterisks indicate that participants were unable to complete all trials for the given placement: participant 2 for prescribed BIC, TRA, and FLE motions; participant 7 for prescribed SCM and TRA, self-selected DEL, one of two self-selected TRA, and one of two prescribed DEL; and participant 5 (uninjured) for one of two prescribed TRA. These nonperformant configurations largely correspond to users' reported deficiencies in function but do not prevent the use of the corresponding sensor configurations altogether, highlighting the potential utility of the system for users with substantial levels of impairment.}
        \vspace{-1em}
        \label{fig:Grid}
    \end{figure}

As demonstrated by the mean RMSE values for each participant and sensor placement documented in Figure~\ref{fig:Grid} (i.e., by the completeness of the table), every participant demonstrated some level of control with every sensor placement. SCI users exhibited slightly, but not meaningfully higher error than most uninjured participants overall, with the placements with highest RMSE and the (few) reported calibration failures generally corresponding to sensor placements where users reported lower levels of residual function. Notably, even when the probe was placed across muscles for which the user had no volitional control, if the user could identify motions generating sufficient tissue deformation (e.g., leveraging residual antagonist function), these sensor placements could still be used. A good example is participant 2, who reported no flexor function and was unable to execute the prescribed wrist flexion/extension motion, but was able to use the FLE sensor placement via a forearm supination/pronation mapping. DEL was particularly non-performant for SCI participants, likely due to the substantial strength required for shoulder abduction and participant-reported interference with trunk balance control (see section~\ref{sec:pref}).

Figure~\ref{fig:Grid} also highlights the extent to which the most performant placement varied substantially by user: all 6 placements had at least one participant for which that placement exhibited the lowest RMSE, and at least one participant was able to achieve $<$4\% error using each placement. This heterogeneity supports the utility of a placement-agnostic system that users can tailor to their individual capabilities and preferences.

The SCM placement is of particular interest due to its potential utility for users with more severe injury and less residual function. Although participants noted challenges with comfortable probe attachment (see section~\ref{sec:pref}), and investigators noted signal noise related to carotid artery pulse, most users were largely able to compensate for such noise and achieve comparable tracking performance to that of other sensor locations.

\subsubsection{Performance Across Motion Speeds \& Patterns}\label{sec:speed} %

\begin{figure}[!t]
        \centering
        \includegraphics[width=0.9\linewidth]{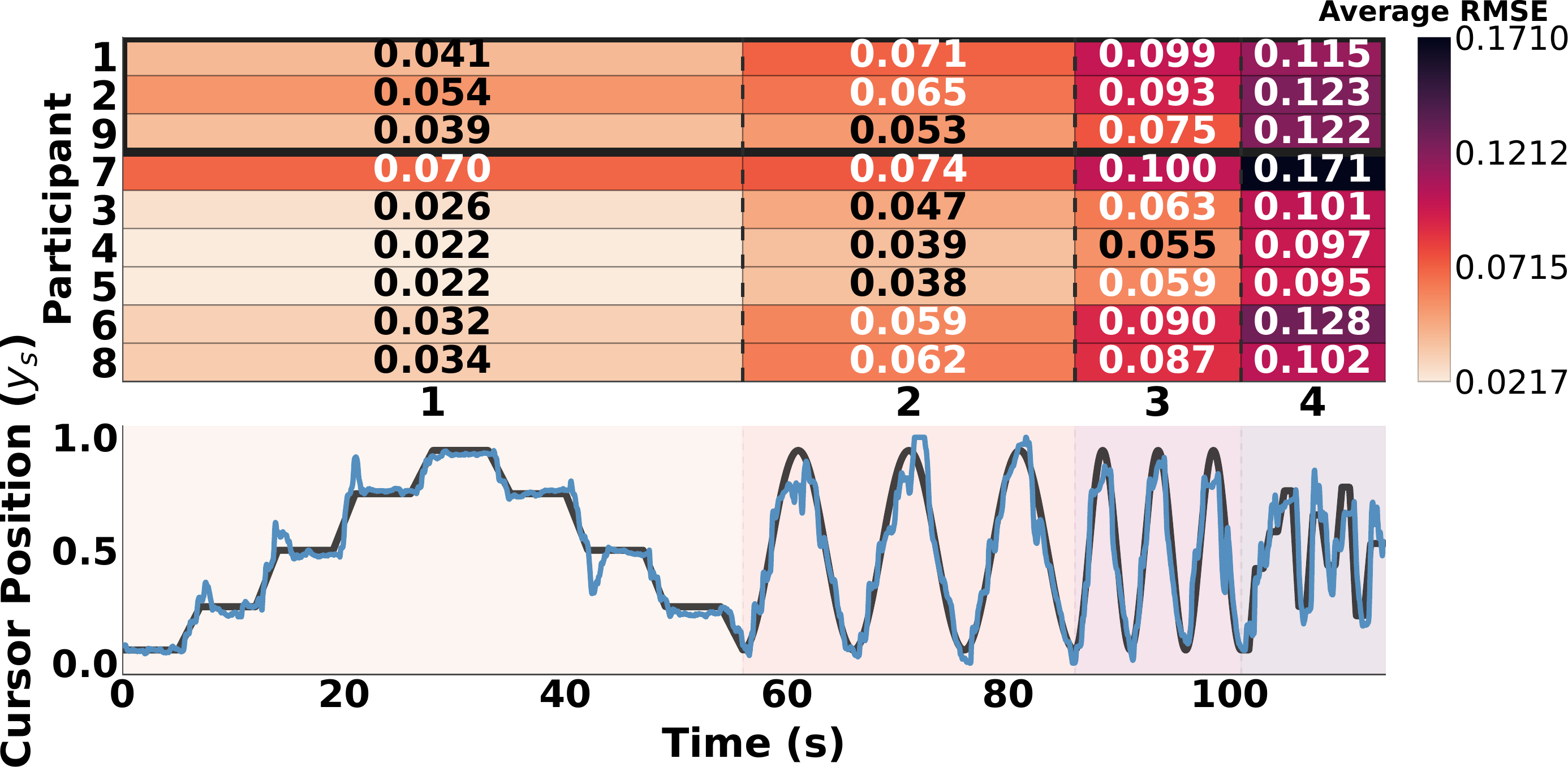 }
        \caption{\textit{Top}: Average RMSE for each participant during each segment of the 1-DOF trajectory tracking task across all successful trials and sensor placements. 
        Early segments highlight the strong performance of the system during slow motions and holds --- a notable advantage over most sEMG-based systems --- while later segments reflect decreasing performance associated with more challenging motions and signal drift. 
        \textit{Bottom}: Exemplar target (\textit{gray}) and user-generated (\textit{blue}) trajectories (participant 2, placement EXT, prescribed motion) illustrate the primary underlying sources of error (deviations at workspace bounds and overshoot when changing direction or stopping), motivating future signal processing refinements.
        }
        \vspace{-1em}
        \label{fig:ErrVsTime}
    \end{figure}

Examining error at each phase of the prescribed trajectory 
yields greater insight into the source of the errors documented in Figure~\ref{fig:Grid}. Figure~\ref{fig:ErrVsTime} (\textit{top}) shows this RMSE stratified across the four different trajectory segments, alongside an exemplar trajectory. Low RMSE on the initial stair-step segment (just over 2\% for some participants) shows how well the system can perform under smooth motions and holds, a capability that is especially valuable given how much processing is generally required to recreate this functionality with more standard modalities like sEMG~\cite{song_review_2023}.

All participants' RMSE increased with each subsequent segment as trajectory speed and difficulty increased, likely exacerbated by point drift in the optical flow algorithm, a known limitation of Lucas--Kanade optical flow tracking~\cite{wong_uncertainty_2017}.
The exemplar trajectory in Figure~\ref{fig:ErrVsTime} (\textit{bottom}) illustrates the underlying source of this error, as overshoot and lag in reaction time happens more often due to more frequent direction changes and users more often visit their workspace extrema, where performance degrades as the optical flow algorithm sometimes erroneously saturates, which could in the future be addressed with appropriate signal filtering and workspace reduction. At the same time, the best-performing users were still able to achieve $\approx$10\% error on average during even the final segment's challenging, highly dynamic task, illustrating the responsiveness of the system.

\subsubsection{Performance Across Motion Magnitudes}\label{sec:motion} %

\begin{figure}[!t]
        \centering
        \includegraphics[width=0.9\linewidth]{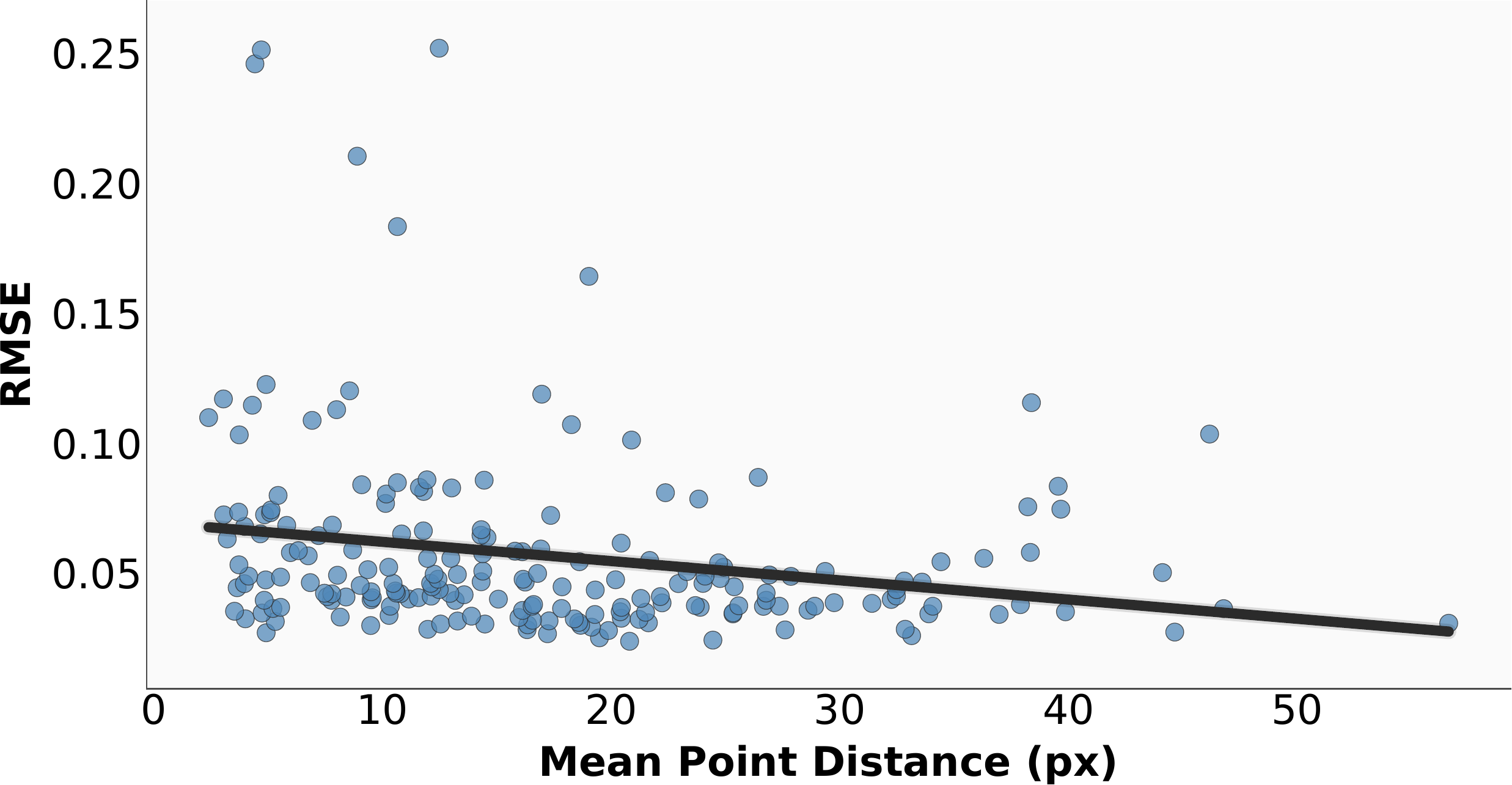}
        \caption{Scatter plot relating mean distance between tracked points in their ``up'' and ``down'' positions to trajectory RMSE for all successful trials, at all sensor placements, for all participants, alongside best-fit linear regression line ($R^2$= $0.0466$, $p$=$0.002$). RMSE is significantly anticorrelated with mean distance --- i.e., larger motions generally achieve better performance --- but many participants achieved comparably low RMSE values with minute changes in pose, supporting the potential utility of the system for individuals with limited motion capabilities.}
        \vspace{-1em}
        \label{fig:pointDisp}
    \end{figure}

Because the magnitude of the driving movements was not specified, users exhibited a wide range of motion sizes, resulting a correspondingly wide range of mean distances between tracked points in their ``up'' and ``down'' positions. The relationship between these distances and control performance is shown in Figure~\ref{fig:pointDisp}, which illustrates that larger mean distances and RMSE are anticorrelated, a finding consistent with qualitative user feedback, as most users expressed having much finer control over trials calibrated with large changes in pose. At the same time, this relationship is relatively flat, and many participants achieved comparably low RMSE with extremely small motions, some with a mean traversable distance of 5~px or less. This is an especially powerful result for the target SCI survivor population, as many potential users are limited to small ranges of motion.

\subsection{2-DOF System Performance} %

\begin{figure}[!t]
        \centering
        \includegraphics[width=0.7\linewidth]{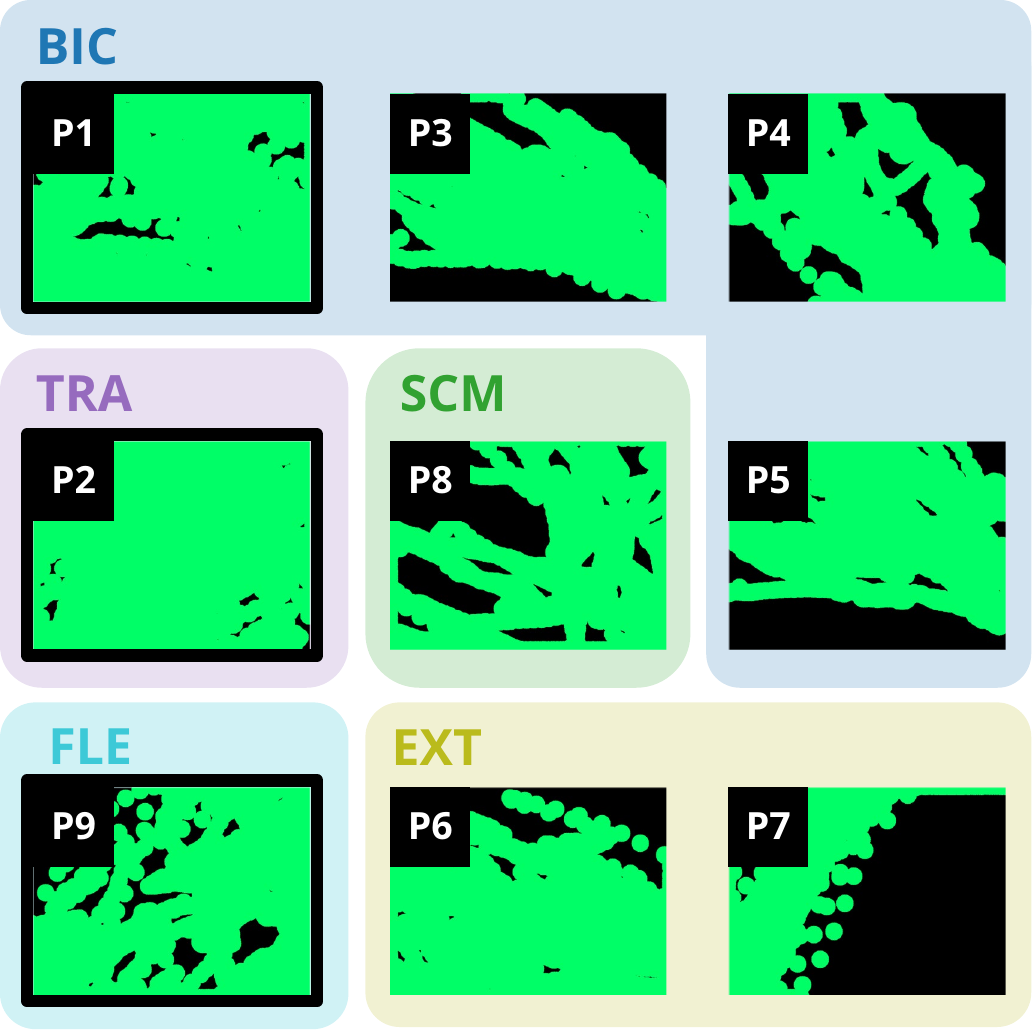}
        \caption{Two-dimensional workspace traversed by each participant during their best (i.e., highest coverage) execution of the 2-DOF workspace traversal task, completed at the noted sensor placements. All participants were able to traverse both workspace dimensions to some extent, but performance varied, with several participants (3, 4, 5) only able to modulate the signals in a coupled manner, resulting in inaccessible quadrants of the workspace, and several others (e.g., 6, 7, 8) experiencing signal ``binding'' along one or more axes, preventing traversal in the orthogonal dimension. Interestingly, SCI survivor participants (\textit{black borders}) exhibited markedly more complete workspace traversal than all uninjured participants, perhaps reflecting these participants' additional requested practice time, which was generally more extensive than that of uninjured participants. 
        Participants' performance across 5 different sensor locations supports the feasibility of system usage by users with varied neuromusculoskeletal function.
        }
        \label{fig:WorkSpcace}
    \end{figure}

Users elected varied sensor placements 
at which to test the 2-DOF system, with specific selections noted in Figure~\ref{fig:WorkSpcace}. Overall, users found the system challenging to use, with nonlinear relationships between input motions and output positions, and for many users, substantial signal coupling. These challenges are expected, as the naive 2-DOF algorithm piloted in this work assumes that orthogonal feature motions are largely independent, which is not the case for the complex soft tissue motions driving the system. The complicated interplay between soft tissue dynamics, model linearity assumptions, and technical challenges (e.g., point drift) make the specific source of particular signal inconsistencies hard to pinpoint, and future research is needed to establish methods to compensate for observed irregularities, which included coupling, ``binding'' to one or more axes, and spatially varying signal sensitivity.

Nevertheless, as shown in Figure~\ref{fig:WorkSpcace}, while this coupling prevented some users from reaching the full 2D cursor workspace, others --- notably, all 3 SCI survivor participants --- were able to adjust to these signal aberrations to traverse the full workspace. Anecdotally, these participants requested longer time to explore the system prior to completing the workspace traversal task, indicating that sufficient training time may allow users to control even unoptimized control schemes like the one tested here.

\begin{figure}[!t]
        \centering
        \includegraphics[width=0.5\linewidth]{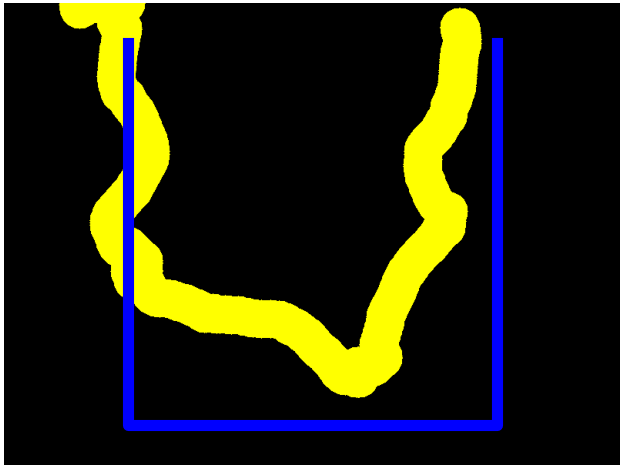}
        \caption{Exemplar target (\textit{blue}) and user-generated (\textit{yellow}) trajectories (participant 8, placement SCM) illustrate the trajectory following capabilities of the current 2-DOF system, the first (to the authors' knowledge) demonstration of continuous, location-agnostic, 2-dimensional SMG-based control. Users found the independent signal modulation required to complete this task challenging, with performance varying widely, illustrating the need for further system refinement to account for signal nonlinearities and interdependencies.
        }
        \vspace{-1em}
        \label{fig:blockU}
    \end{figure}

In addition to workspace traversal, some users were able to complete the stylized letter ``U'' drawing task using the same control mapping; the most successful execution across all participants is shown in Figure~\ref{fig:blockU}. While the signal shows signs of the nonlinearities described above, this is, to our knowledge, the first demonstration of location-agnostic sonomyographic control in 2 continuous, independently-modulated dimensions, a key milestone toward SMG interfaces that can drive multi-DOF systems. Notably, the depicted participant accomplished this task using the SCM, supporting feasibility of such a system by even users with high-level injuries.

\subsection{User Preferences}\label{sec:pref} %

We focus our subjective analyses on survey data from the 3 SCI survivor participants, as their impressions are most critical for assessing the usability and potential impact of our developed interface. Overall, these participants were optimistic about the potential use cases for such a system, with all 3 ranking it as ``very useful'' (5/5) when asked about its suitability for in-home robotic assistance. Participants' qualitative comments supported this excitement (as did their uniformly high SAM valence scores), with many describing the control schemes as ``responsive,'' ``entertaining,'' ``super fun,'' and having ``a lot of potential.'' Participants also noted some important areas for system improvement --- most notably, limitations of the probe geometry and mounting systems, which many users in both SCI and uninjured cohorts found uncomfortable (though all were able to tolerate at least one attachment method for the duration of data collection). This is a challenging, but important limitation to address in future system development, as flexible B-mode ultrasound probes are not commercially available, and the length and rigidity of our current setup's transducers interface poorly with many areas of the body, motivating the need for future hardware development to enable a truly sensor-placement-agnostic system. Designing a placement-agnostic attachment mechanism is similarly challenging due to the highly variable geometry of the human body, and this is one area in which we may design location-specific system elements in future iterations.

    \begin{figure}[!t]
        \centering
        \includegraphics[width=1\linewidth]{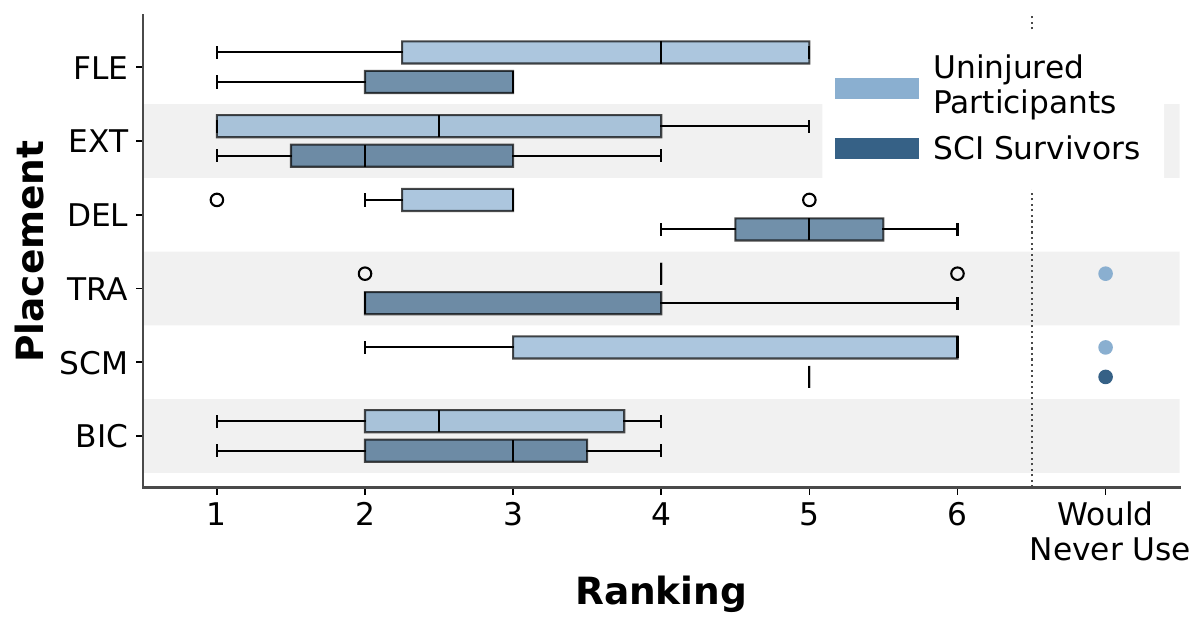}
        \caption{
        Ranking of each sensor placement from 1 (most preferred) to 6 (least preferred), with the option to indicate never wanting to use a given placement, stratified by uninjured (\textit{light gray}) and SCI survivor (\textit{dark gray}) cohorts.
        Most-preferred sensor placements varied across users, with reported preferences for intuitive, non-fatiguing control that sometimes but did not always correspond to actual control performance as quantified in Figure~\ref{fig:Grid}. Preferences across cohorts were not obviously different, except that SCI participants showed a markedly lower preference for DEL, reflecting both degraded performance and reported challenges maintaining trunk balance while performing this task. Participant 7 reported never wanting to use SCM or TRA placements due to their nonperformance (likely due to that participant's adhesive capsulitis), and participant 2 reported never wanting to use the SCM placement due to discomfort with the probe mounting system.
        }
        \vspace{-2em}
        \label{fig:participantPref}
    \end{figure}

When asked to rank sensor placements in order of preference, users' rankings were heterogeneous, as illustrated in Figure~\ref{fig:participantPref}, which notes both discrepancies and commonalities across uninjured and SCI user preferences. The most preferred locations were the bicep (BIC) and extensor (EXT), with 4 participants ranking BIC within their top two preferences, and 4 ranking EXT as their first preference. 
Many participants communicated prioritizing intuitive control, minimal fatigue, and for those with SCI, control schemes that did not interfere with trunk balance. Participants highlighted control schemes for which motions of the body and the on-screen cursor were kinematically similar (e.g., BIC/EXT/FLE motions at which upward cursor motion corresponded to flexion at the elbow or wrist) as particularly easy to control. In addition, one participant enjoyed the EXT placement due to its novelty, interjecting that ``the whole Jedi thing really got in my head, this is amazing,'' suggesting that in addition to comfort and controllability, how entertaining the system is to use plays a role in user preference that may account for discrepancies across performance and preference.

Although many users ranked the SCM low in preference, one participant --- participant 8, whose successful execution of the 2-DOF drawing task is depicted in Figure~\ref{fig:blockU} --- found this placement especially functional because of their ability to isolate the different motions of the two axes. Furthermore, the control scheme employed was intuitive for 2-DOF control, as neck flexion/extension corresponded nicely with vertical motion and lateral flexion/extension with horizontal motion.

\section{Conclusions, Limitations, \& Future Opportunities}\label{sec:conc}

The strong performance of both SCI survivors and uninjured participants using the 1-DOF continuous control system, and the overall positive impressions expressed by those users, 
are evidence that SMG-based control is a strong candidate for expanded development. Users' heterogeneous performance and preferences across sensor locations illustrate the importance of location-agnostic sensor development in particular, and the ability of users with tetraplegia to control signals with non-innervated muscles as long as other tissue-deformation-generating muscles were intact supports the idea that SCI users may retain more target locations from which to extract control signals than are generally explored. This exploration of many sensor locations was only practical due to the minimal, 3-point calibration required by our system, supporting that in addition to the intrinsic value of non-onerous calibrations, such systems can permit important ``broad'' investigations in the biosensing space rather than the ``deep'' collections required by data-hungry models that are currently in vogue.

This preliminary evaluation of our novel 1- and 2-DOF systems was limited in scope, with many opportunities for future development. All 3 SCI survivor participants retained some volitional arm control, and the proposed control schemes could be especially beneficial to those with no such function. While several examined sensor positions (SCM and TRA) would likely be viable for such participants, we aim to validate this assumption with expanded recruitment of participants with varying injury levels and AIS scores. Our study was also limited temporally, with 2~min tracking tasks for each calibration, and did not address questions of signal robustness over the longer time horizons required for practical deployment, nor were precise calibration and training times quantified.

Even among our current participant cohort, many questions remain regarding the mechanisms underlying control performance, especially for the 2-DOF system; while users were able to modulate both signals with varying levels of independence to traverse the workspace, few were able to perform ``useful'' drawing movements with any level of precision, and we do not yet know what differentiates those participants' control strategies from those that were unsuccessful. A rigorous characterization of the interplay between mathematical assumptions, tissue characteristics, selected motions, and users' adaptation strategies is a compelling area for future investigation.

Beyond these immediate shortcomings, our investigations highlighted many opportunities and challenges for future development. In particular, we aim to refine our formulation of the 2-DOF system to address observed signal aberrations (both through expanded post-processing and novel methods of signal decomposition), and to expand our controllers to include multiple probes (enabling higher-DOF outputs), sEMG components (e.g., to address optical flow signal drift), and output systems more representative of target use cases (virtual and physical robots). We also aim to explore these systems over longer learning and adaptation horizons, which can enable substantial performance improvements~\cite{lee2024learning}.

The problem of continuous high-DOF device control is multifaceted, and will require addressing not only the signal processing algorithms that were the focus of this work, but improvements in hardware design of both the sensor and associated interface. We're excited to open-source
our algorithms 
through this paper's associated data release, and look forward to the community's collective advancements enabling improved device control and user empowerment.

\section*{Acknowledgment}

The authors acknowledge the contributions of Aaron and Amanda Collyer, Denise Gregg, Jennifer Molnar, Ari Sanders, Gabriel Parra, Simon Padgen, and Ajay Anand.

\bibliographystyle{IEEEtran}
\bibliography{ref.bib}
\end{document}